\documentstyle[12pt,emulateapj]{article}
\input{epsf}
\journalid{}{}
\articleid{}{}

\begin{document}
\def\lax    {\ifmmode{_<\atop^{\sim}}\else{${_<\atop^{\sim}}$}\fi}
\def\gax    {\ifmmode{_>\atop^{\sim}}\else{${_>\atop^{\sim}}$}\fi}
\def\gtorder{\mathrel{\raise.3ex\hbox{$>$}\mkern-14mu
             \lower0.6ex\hbox{$\sim$}}}
\def\ltorder{\mathrel{\raise.3ex\hbox{$<$}\mkern-14mu
             \lower0.6ex\hbox{$\sim$}}}
 
\long\def\***#1{{\sc #1}}

\title{OBSERVATIONS OF THE X-RAY AFTERGLOWS OF GRB~011211 AND GRB~001025 
BY XMM-NEWTON}
\author{Konstantin N.~Borozdin, 
 Sergey P.~Trudolyubov
}
\affil{NIS-2, Los Alamos National Laboratory, Los Alamos, NM 87545, U.S.A.}

\begin{abstract}
We present the {\it XMM-Newton} observations of X-ray afterglows of
the $\gamma$-ray bursts GRB~011211 and GRB~001025. 
For GRB~011211 {\it XMM} detected fading X-ray object with an average flux in 
0.2-10~keV declining from $2.7\times10^{-13}$~erg~cm$^{-2}$~s$^{-1}$ during 
the first 5~ks of 27-ks observation to $1.0\times10^{-13}$
~erg~cm$^{-2}$~s$^{-1}$ toward the end of the observation.  The spectrum
of the afterglow can be fit to a power law with $\alpha$=2.16$\pm$0.03 
modified for the Galactic absorption. No significant evolution 
of spectral parameters has been
detected during the observation. Similar X-ray spectrum with 
$\alpha$=2.01$\pm$0.09 has been observed by the XMM 
from the GRB~001025.  The non-detection of any extra absorption in these
spectra above the Galactic value is an interesting fact and may impose 
restrictions to the favorable GRB models involving burst origin in
star-forming regions. Finally we discuss soft X-ray lines from
GRB~011211 reported by Reeves et al. and conclude
that there is no definitive evidence for the presense of these lines 
in the spectrum.
\end{abstract}



\section{Introduction}
\label{secintro}

Cosmic bursts of gamma-rays are some of the most elusive and mysterious 
events in the Universe. Typical burst lasts less than a minute and disappears 
thereafter. However, since the {\it BeppoSAX}'s first discovery 
on 1997 February 28 (Costa et al. 1997)
it is known that gamma-ray bursts (GRBs) also produce 
X-ray afterglows, which can be detected with modern X-ray observatories 
for several hours and sometimes several days after the burst. 
More than a few X-ray afterglows 
have been observed in last five years by {\it BeppoSAX, ASCA and Chandra} 
satellites.  Several afterglow observations were carried out with 
{\it XMM-Newton}.  In this {\it Letter} we report on two of the latter
observations.

GRB 011211 was detected and localized in {\it BeppoSAX} WFC1 on 
2001 December 11, 19:09:21 UT(Gandolfi 2001).
The distinguishing features of GRB~011211 were its long duration (the
longest event localized with {\it BeppoSAX}) and its faintness 
both in X- and $\gamma$-rays (Frontera et al. 2002).
Grav et al. (2001) 
discovered a new point
source in optical R-band (R$\simeq$19) at 
R.A.=11$^h$15$^m$17$^s$.98, 
Dec=$-$21$\arcdeg$56$\arcmin$56$\arcsec$.2 (J2000, $\pm$1$\arcsec$)
identified as the afterglow of GRB~011211. 
Fruchter et al. (2001) and Gladders et al. (2001)
measured the optical spectrum 
and found an absorption line system corresponding to a 
redshift of z=2.14. 
Later the optical transient was found to be superposed 
on an apparent host galaxy with R = 25.0$\pm$0.3 (Burud et al. 2001).

GRB~001025 was detected by 
the {\it RXTE All-Sky Monitor} on 2000 October 25 at about 03:10:05~UT 
(Smith et al. 2000).  
lasted approximately 15-20~s
and reached a peak 5-12~keV flux of $\sim$4 Crab.
Optical afterglow of GRB~001025 has never been detected
down to R=24.5 for any significantly variable object 
(Fynbo et al. 2000).

\section{XMM-Newton Observations and Data Analysis}
\label{sectionburst}

The observation of GRB~011211 afterglow 
was started on 2001 December 12 at 06:16:56~UT (Santos-Lleo et al. 2001).
A source was initially located 
so close to the edge of CCD\#7 in EPIC-PN detector that some of 
the source photons were falling in the inter-CCD gap. 
Therefore, the telescope was re-pointed 
The re-pointing slew started at 08:31:16UT and finished at 08:40:50UT. 
Useful exposure of total observation is $\sim$27~ks.

XMM-Newton observed the location of GRB~001025
from Oct.27.003 to Oct.27.46~UT, starting $\sim$1.9 days after the burst.
Two X-ray sources were detected in the error box (Altieri et al. 2000).
Below we discuss the brighter source (R.A.=8$^h$36$^m$35$^s$.92,   
Decl.=-13$\arcdeg$04$\arcmin$09$\arcsec$.9, J2000).

We analyzed data products from the two {\em XMM-Newton} observations. 
In all observations the EPIC instruments (Turner et al. 2001, 
Strueder et al. 2001) were
operated in the {\em full window mode} ($30\arcmin$ diameter FOV). The 
medium optical blocking filter was used with MOS detectors for GRB~011211
observation.  Thin filter was used with PN detector for GRB~011211 and
with MOS detectors in GRB~001025 observation. No PN data were available for
GRB~001025. We reduced EPIC data with the {\em XMM-Newton} Science Analysis 
System (SAS v.5.2)\footnote{See http://xmm.vilspa.esa.es/user}. 
EPIC detectors data were fit together with common model
parameters. The count rates were converted into energy fluxes using analytical 
fits to the spectra. We have used Galactic absorption values provided 
by the HEASARC\footnote{http://heasarc.gsfc.nasa.gov} ``nH'' tool 
(Dickey \& Lockman, 1990).

\section{Results}

\subsection{GRB~011211}

Central part of MOS1 detector (converted to celestial coordinates)
is shown in {Fig.\ 1a}. The bright source in the middle of the image
corresponds to the position of GRB~011211 optical afterglow.
Significant X-ray flux decline during the observation (Fig.\ 1b)
provides an additional evidence for an identification of the X-ray source
with the GRB.  X-ray light curve can be fit with F$_X \sim$t$^{-\beta}$
($\beta$=1.5-1.7) or with F$_X \sim$e$^{-t/\tau}$ ($\tau$=30.$\pm$0.5~ks).
Average measured flux in 0.2-10~keV band was equal to 
1.7$\times$10$^{-13}$ erg/s/cm$^2$. Extrapolating back to t=100~s since
the initial detection of the burst we get an X-ray flux 
F$_X$=5$\times$10$^{-9}$ erg/s/cm$^2$ 
or X-ray afterglow luminosity L$_X$=2$\times$10$^{50}$
erg/s (0.6-30 keV) in the rest frame at z=2.14 and a
$H_0$=65~km~s$^{-1}$Mpc$^{-1}$, $\Omega_{\rm m}=0.3$, $\Lambda=0.7$ 
cosmology. Spectrum of the GRB~011211 afterglow can be described 
as a simple power-law with photon index $\alpha$=2.16$\pm$0.03 
modified by the Galactic absorption only (Fig.\ 2a).  
No significant spectral evolution has been found
in a sequence of 5-ks intervals within the observation (Fig.\ 2b).

\subsection{GRB~001025}

We built the spectrum of the brightest source in error box 
of GRB~001025 (Fig.\ 3). Combined spectrum of two MOS detectors can be readily
approximated by a power-law with photon index $\alpha$=2.01$\pm$0.09 
and Galactic absorption value N$_H$=6$\times$10$^{20}$~cm$^{-2}$.
Average flux from the source during the observation 
was 5.3$\times$10$^{-14}$erg/s/cm$^2$.
No significant flux variability was detected, but the spectrum
is very typical for X-ray afterglows of GRB, supporting
the identification of this source with GRB~001025.

It is interesting to consider an alternative identification
of the X-ray source with a background quasar.  An X-ray bright 
quasar 3C273 has a R magnitude of $\sim$12.5 (Odell et al. 1978)
and the 0.5-10~keV X-ray flux of 23$\times$10$^{-11}$ erg/cm$^2$/s
(Reeves \& Turner 2000).
Neglecting the K-correction and simply scaling the R magnitude to the 
X-ray flux we find that if a 3C273-like quasar is a source of
the detected X-ray flux of $\sim$4.7$\times$10$^{-14}$ erg/cm$^2$/s
(0.5-10~keV) then its optical counterpart would have a R 
magnitude of $\sim$21.7.  As in fact an optical counterpart for
GRB~001025 had not been found down to R=24.5 (Fynbo et al. 2000) 
we consider unlikely that the X-ray source is an AGN.
\section{Discussion}

\label{sectiondis}
\subsection{X-ray spectra of GRB afterglows}

We have detected power-law spectra with index $\sim$2 from two GRB
afterglows.  In case of GRB~011211
an identification of the X-ray source with the GRB is supported by
the observation of optical transient and also by the decline of X-ray
flux during the {\it XMM-Newton} observation.  
For GRB~001025 we do not have such supporting evidence,
but the spectral shape itself together with the position of X-ray source 
inside IPN/RXTE error box allows us to suggest that the source 
is indeed an afterglow of GRB~001025. 
Power-law fit with slope $\sim$2 is very typical for detected X-ray
afterglows of GRBs (see e.g. Harrison et al. 2001, in't Zand et al. 2001, 
Antonelli et al. 2000, ).  We did
not detect any significant changes in the X-ray spectrum during
the long observation of GRB~011211.  Afterglow of GRB~001025 was 
observed significantly later after the burst, and measured 
X-ray flux was much lower, 
but the spectrum was almost identical to GRB~011211.
Conspicuously the X-ray spectra in all or most of the observed
afterglows are generated by common physical process 
and do not depend much on the differences in 
the burst environments. 
Overall spectral shape can be fit to popular model of synchrotron
emission with possible inclusion of inverse Compton scattering (Piran 1999; 
Granot \& Sari 2002; Sari \& Esin 2001).

No significant absorption above the
Galactic values has been detected in the X-ray spectra.  
High absorption would be naturally expected if the burst occurs in 
a high-density star-forming region.  Ramirez-Ruiz, 
Trentham \& Blain (2002) suggested that high absorption in nearby
(relative to the burst birthplace)
interstellar media may be the reason for a lack of optical 
detections in a significant fraction of GRB. 
The absorption should be detectable in soft X-rays (0.5-2~keV).  
Contrary to such expectations, we did not detect any significant 
absorption above the Galactic value in both a GRB with optical afterglow
(GRB~011211) and without it (GRB~001025). 
The lack of X-ray absorption in GRB~011211 is consistent with the lack
of detected reddenning in the spectrum of optical afterglow
(Simon et al. 2001).

\subsection{On the spectral lines in GRB~011211}

Reeves et al. (2002a, hereafter RWO) reported on the discovery 
of a blue-shifted line complex in the spectrum of GRB~011211.  
The lines have been detected only with PN detector and only during 
first 5~ks of the observation, with the source located close to the CCD 
chip boundary (Fig. 4a).  We have extracted source and background
spectra from the same regions as RWO (Reeves et al. 2002b).  
The spectrum can be satisfactory fit
to power-law with $\alpha$=2.4 and N$_H$ = 8 $\times$ 10$^{20}$ cm$^{-2}$
($\chi^2$=1.03 for 44 d.o.f., Fig.~4b).  Joint fit of PN, MOS1 and MOS2 data
gives $\chi^2$=0.92 (72 d.o.f.) for $\alpha$=2.15$\pm$0.06 and an absorption
fixed at the Galactic value.  Hence we are able to get the perfect fit
using a simple power-law model we do not see much value in adding
extra lines to the model.  Addition of 5 lines at the energies 
specified by RWO gives $\chi^2$=0.93 (62 d.o.f.) for joint fit 
of the PN and MOS data.  The analysis of PN data alone allowed
us to reproduce line fluxes reported by RWO with somewhat lower significance
(Fig. 4c and Table 2).  We note that the PN data should be treated 
with special caution for the first 5~ks of the observation because 
of the unfortunate position of the source on the CCD.  In their 
analysis RWO collected source data from two different PN chips and from 
the areas near the interchip boundary, which are the least suitable
for fine spectroscopy.  Our extensive analysis showed that any alternative
choice of extraction regions for the source and background leads to
further reduction in the lines significance.  We also found that 
background spectrum collected over the interchip edge is dominated by
a strong feature at 0.7 keV, exactly at the energy of the most significant
``line'' reported by RWO (see Fig.4f).  Though the bulk of the events
forming this line are eliminated by the event filtering we are still
concerned that some of these bad events may be present in the spectrum
of GRB~011211. Our concern is amplified by the lack of any of the RWO 
``lines'' in the PN spectrum after the reorientation (Fig.4d). 
Spectral evolution of the source in sync with
satellite revolutions looks quite suspicious unless one suggests
that presense of the ``lines'' depends on the position of the source on 
the chip.  Such alternative hypothesis is further confirmed 
by a lack of ``lines'' in the MOS data (Fig. 4e).
It may therefore be concluded that there 
is no definitive evidence for the presence of the line complex 
at a redshift of 1.88 in the X-ray spectrum of GRB~011211. 
The existence of such complex is a possibility,
but its statistical significance is greatly overestimated by RWO.
Our analysis suggests that the spectrum of GRB~011211 is featureless
and does not contain any significant line emission.

\acknowledgements

We have used publicly available data obtained with {\em XMM-Newton} 
satellite. 
We are grateful to the personnel of the {\em XMM-Newton} Science 
Operations Centre at VILSPA, Spain for satellite 
operations and expedited preparation of data products 
for scientific analysis. We are thankful to W.Priedhorsky and L.Titarchuk 
for encouraging advices, and to anonymous referee for her/his  
helpful comments.  We acknowledge an interesting discussion 
of the draft version of our paper with J.Osborne.


\begin{table}[t]
\caption[]{Spectral parameters of the GRB~011211 afterglow.\label{table3}}
\begin{tabular}{cccc}
\hline\hline
Time, ks$^a$ & Photon index & $\chi ^2$(d.o.f) & Flux (0.2-10~keV) \\
& & &$\times$10$^{-13}$erg~cm$^{-2}$s$^{-1}$\\
\hline
$<$ 5 & 2.14$\pm$0.06 & 0.91 (74) & 2.69\\
5--10 & 2.17$\pm$0.07 & 1.23 (62) & 2.10\\
10--15 & 2.23$\pm$0.08 & 0.98 (45) & 1.55\\
15--20 & 2.09$\pm$0.08 & 1.11 (41) & 1.57\\
20--25 & 2.06$\pm$0.10 & 0.89 (35) & 1.37\\
$>$ 25 & 2.15$\pm$0.12 & 1.06 (28) & 1.01\\
\hline\hline
\end{tabular}
\begin{list}{}{}
\item[$a$] -- from the start of EPIC-PN observation
\end{list}
\end{table}

\begin{table}[t]
\caption[]{Parameters of the ``emission lines''$^a$ in the spectrum of GRB~011211
(first 5 ks of the observation)
\label{table3}}
\begin{tabular}{ccccc}
\hline\hline
Line& Central & Flux/Significance$^c$ & Flux/Significance$^c$ & Redshift\\
ID$^b$& energy$^b$ & in the PN data & in the PN + MOS data & Z$^b$\\
& keV &$\times$10$^{-15}$erg~cm$^{-2}$s$^{-1}$& $\times$10$^{-15}$erg~cm$^{-2}$s$^{-1}$&\\
\hline
Si XIV & 0.70$\pm$0.02 & 11. / 3.0 $\sigma$ & 6.5 / 2.1~$\sigma$ & 1.86$^{+0.08}_{-0.08}$\\
S XVI & 0.89$\pm$0.01 & 10. / 3.1 $\sigma$ & 7.2 / 2.8~$\sigma$ & 1.94$^{+0.04}_{-0.03}$\\
Ar XVIII & 1.21$\pm$0.02 & 9.3 / 2.8 $\sigma$ & 2.9 / 1.2~$\sigma$ & 1.74$^{+0.05}_{-0.04}$\\
\hline\hline
\end{tabular}
\begin{list}{}{}
\item[$a$] -- data for three most significant lines are provided; the other two lines are less significant (Reeves et al. 2002a, b)
\item[$b$] -- adopted from Reeves et al. (2002a)
\item[$c$] -- with PN regions for source and background as in Revees et al. (2002b) 
\end{list}
\end{table}

\begin{figure}
\vbox{
\hbox{
\begin{minipage}{8.5cm}
\epsfxsize=8.0cm
\epsffile{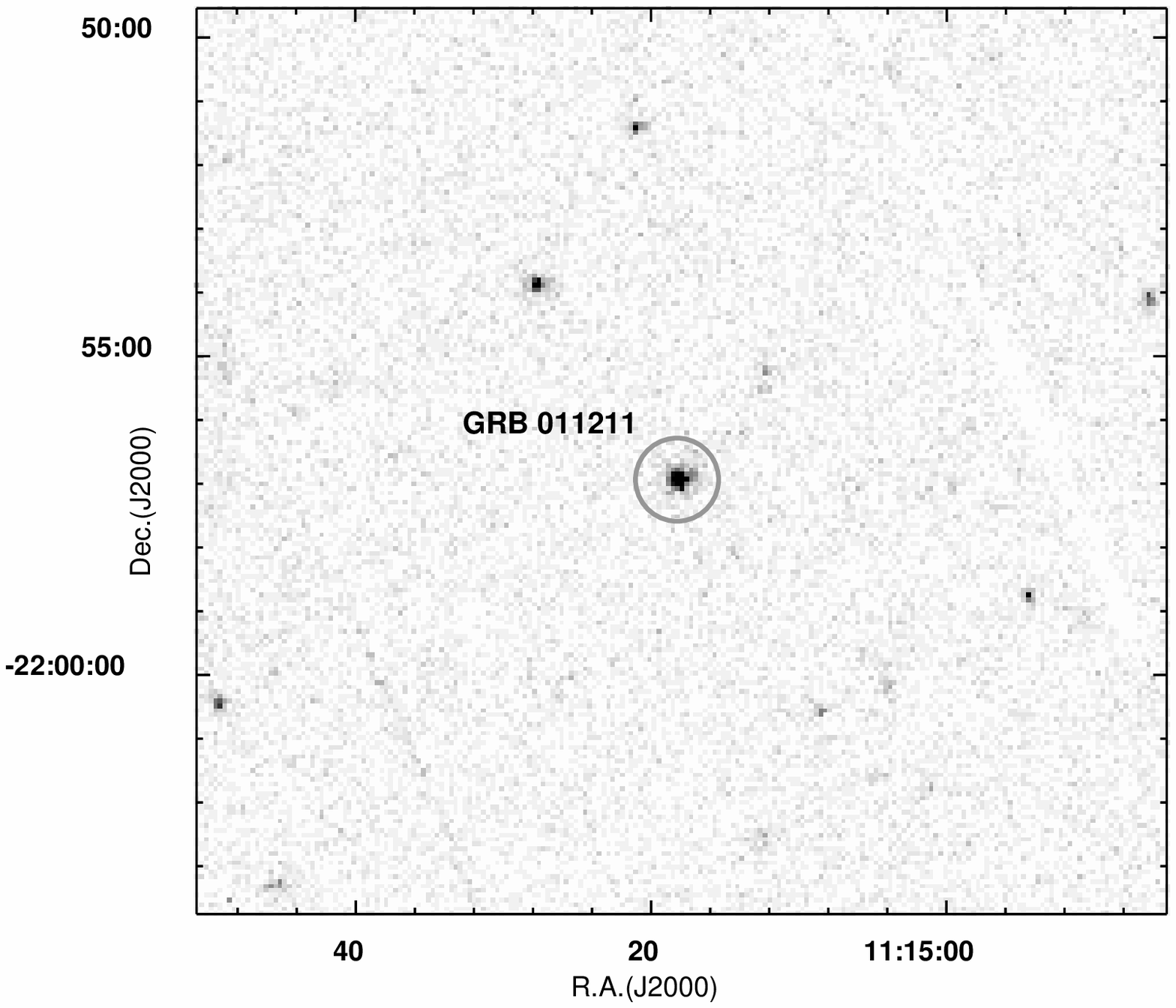}
\end{minipage}
\begin{minipage}{8.5cm}
\epsfxsize=8.0cm
\epsffile{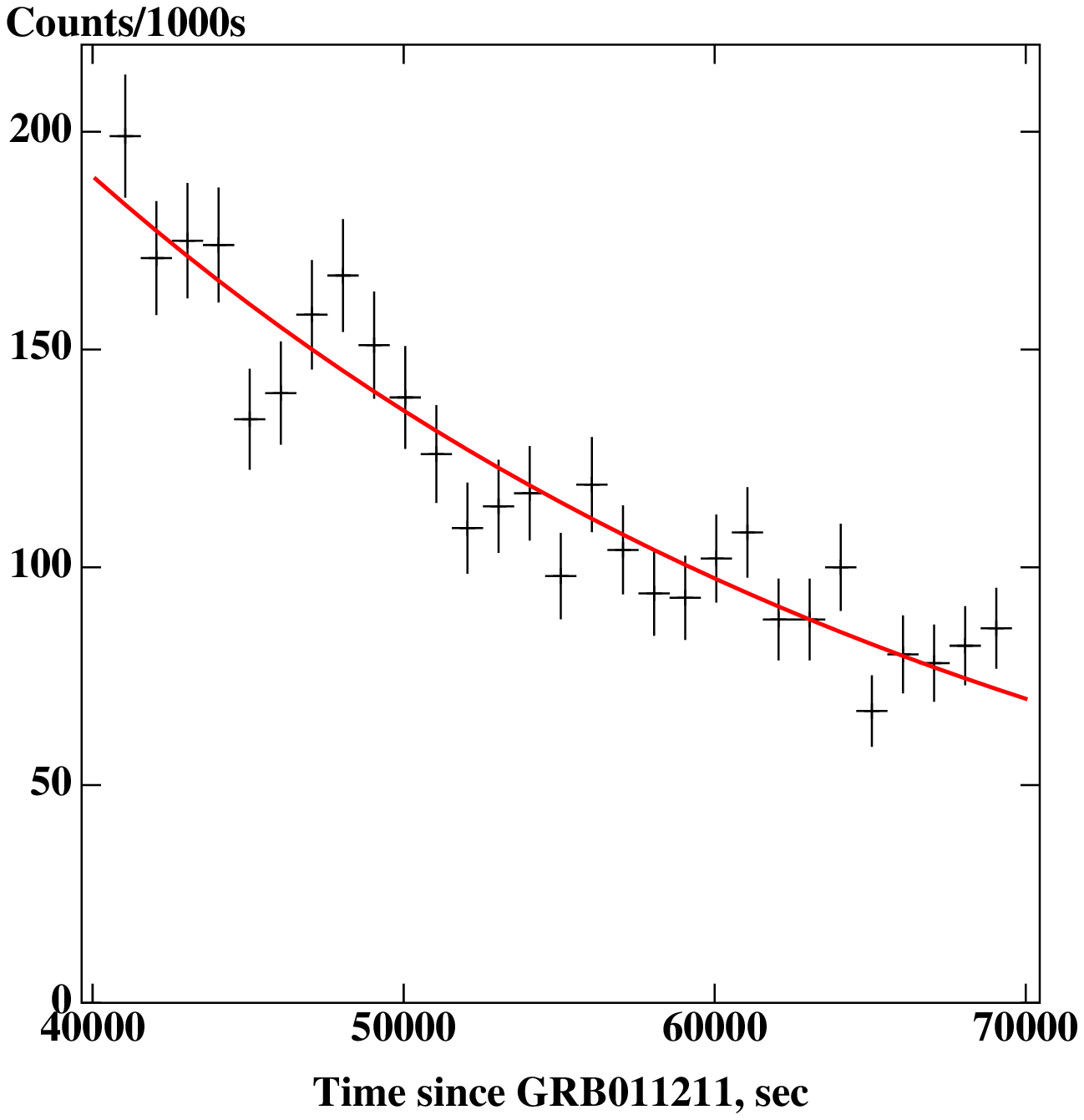}
\end{minipage}
}
}
\caption{EPIC-MOS1 image {\it (left panel)} and 
X-ray light curve {\it (right panel)} of GRB~011211 afterglow 
as observed with XMM-Newton.  Declining light curve is an evidence
for identification of X-ray source with the GRB~011211 afterglow.
Fading of the X-ray flux in 0.2-5 keV band is well described 
with exponential decay F$_X \sim$exp(-t/$\tau$) with $\tau$=30,000$\pm$500~s
(represented by solid line).
\label{GRB01}
}
\end{figure}

\begin{figure}
\vbox{
\hbox{
\begin{minipage}{6.2cm}
\epsfxsize=6.7cm
\epsffile{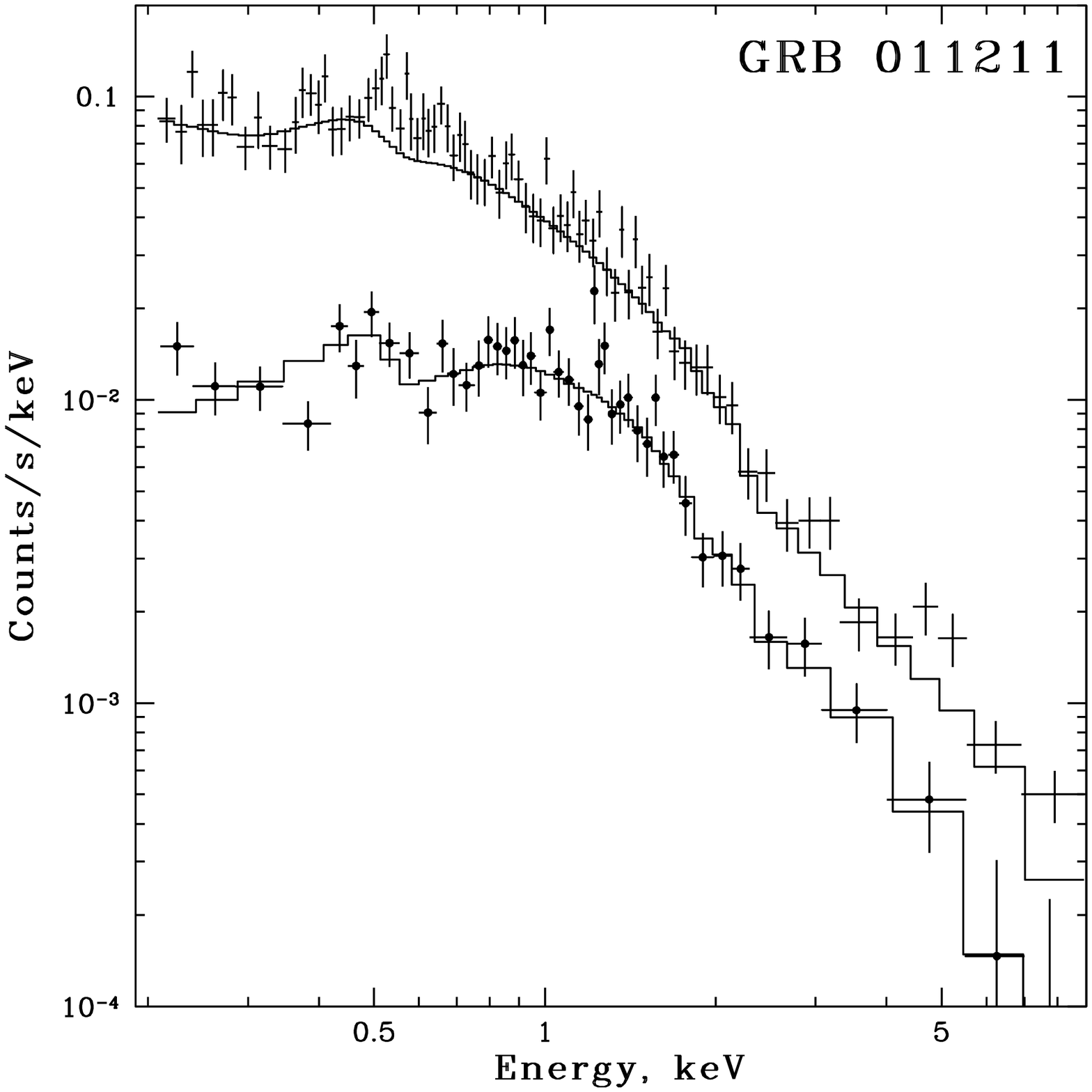}
\end{minipage}
\begin{minipage}{8.0cm}
\epsfxsize=10.7cm
\epsffile{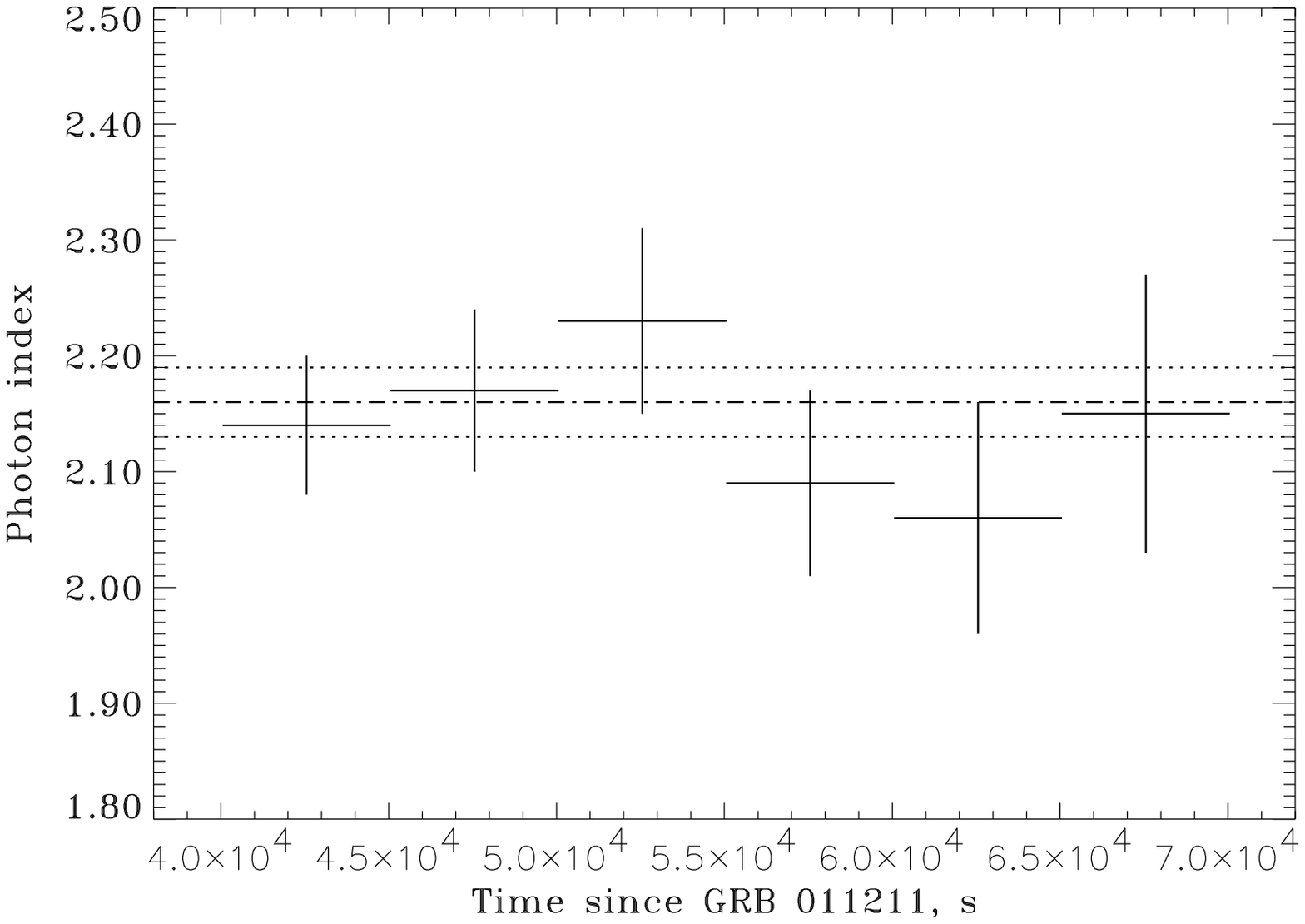}
\end{minipage}
}
}
\caption{ 
X-ray spectrum ({\it left panel}) of GRB~011211 afterglow 
as observed with XMM-Newton. 
Integral spectrum of GRB~011211 afterglows is well described 
with simple power-law modified only by Galactic absorption.
Data from PN ({\it solid crosses}) and  MOS({\it crosses with points})
EPIC detectors were fit jointly to power-law ({\it solid curve}) 
with photon index $\alpha$=2.16$\pm$0.03 
and absorption fixed at the Galactic value 
(N$_{HL}$=4.2$\times$10$^{20}$~cm$^{-2}$). Reduced 
$\chi^2$=0.98 for 235 d.o.f.
No significant evolution of photon index has been detected during 
the observation ({\it right panel}). {\it Crosses} represent best-fit values
of individual spectra, {\it horisontal lines} shows the best-fit
value and error for photon index of integral spectrum. 
}\label{images}
\end{figure}

\begin{figure}
\vbox{
\hbox{
\begin{minipage}{8.5cm}
\epsfxsize=10.5cm
\epsffile{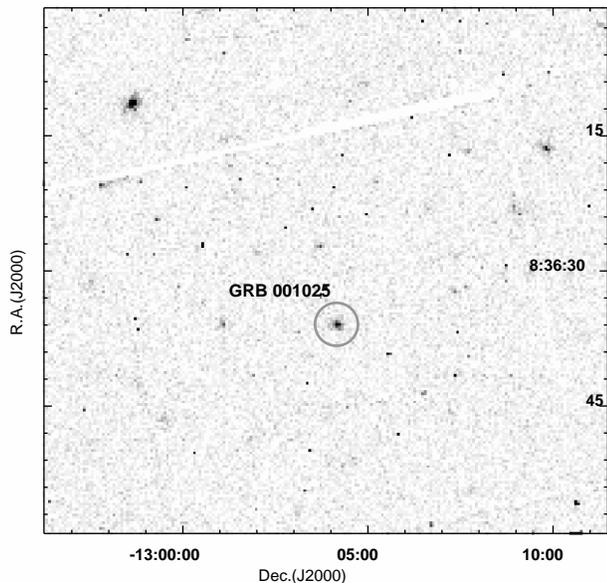}
\end{minipage}
\begin{minipage}{8.5cm}
\epsfxsize=7.7cm
\epsffile{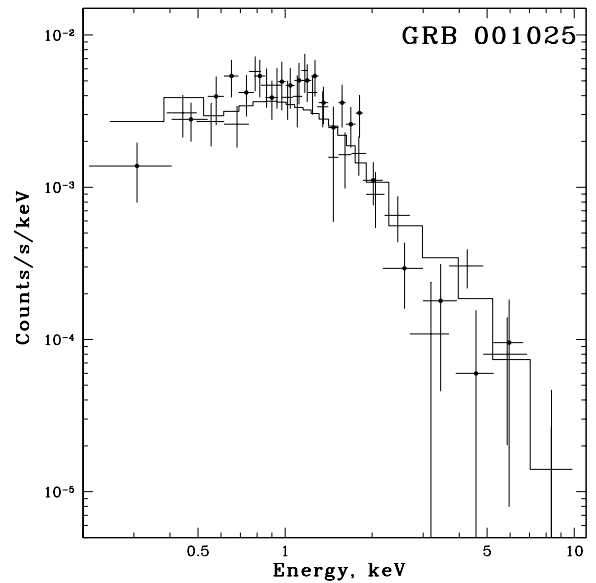}
\end{minipage}
}
}
\caption{Image of MOS1 detector shows the presense of rather bright
X-ray source in the center of field-of-view ({\it left panel}). 
Integral spectrum of GRB~001025 afterglow is well described 
with power-law with the Galactic absorption ({\it right panel}).
Data MOS1({\it crosses})
and MOS2 ({\it crosses with points}) EPIC detectors were fit
together to power-law with photon index $\alpha$=2.01$\pm$0.09 
and absorption fixed at the Galactic value 
(N$_{HL}$=6.0$\times$10$^{20}$~cm$^{-2}$). Reduced 
$\chi^2$=1.18 for 40 d.o.f.
}\label{images}
\end{figure}

\begin{figure}
\vbox{
\hbox{
\begin{minipage}{5.7cm}
\epsfxsize=6.0cm
\epsffile{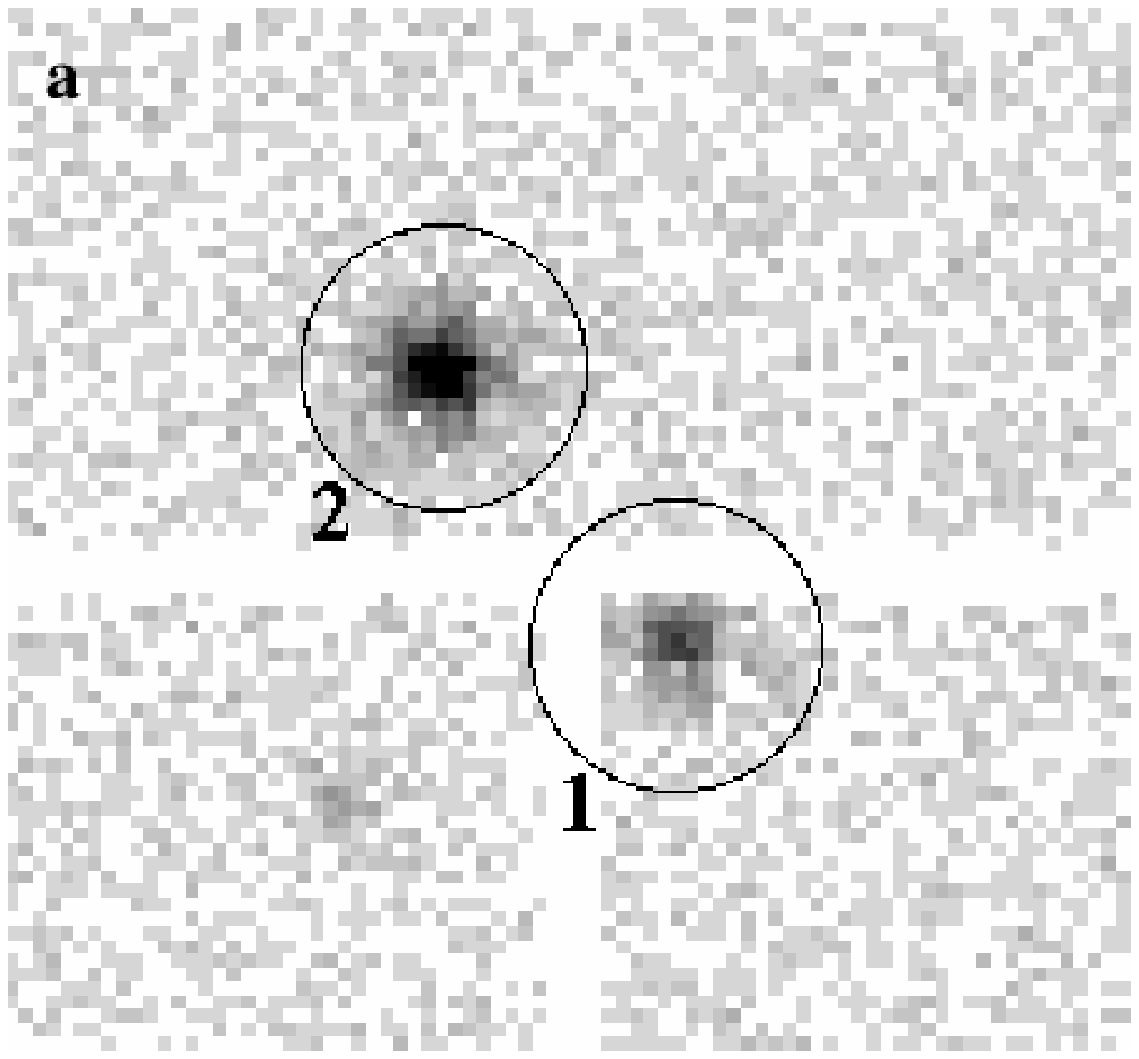}
\end{minipage}
\begin{minipage}{5.7cm}
\epsfxsize=5.9cm
\epsffile{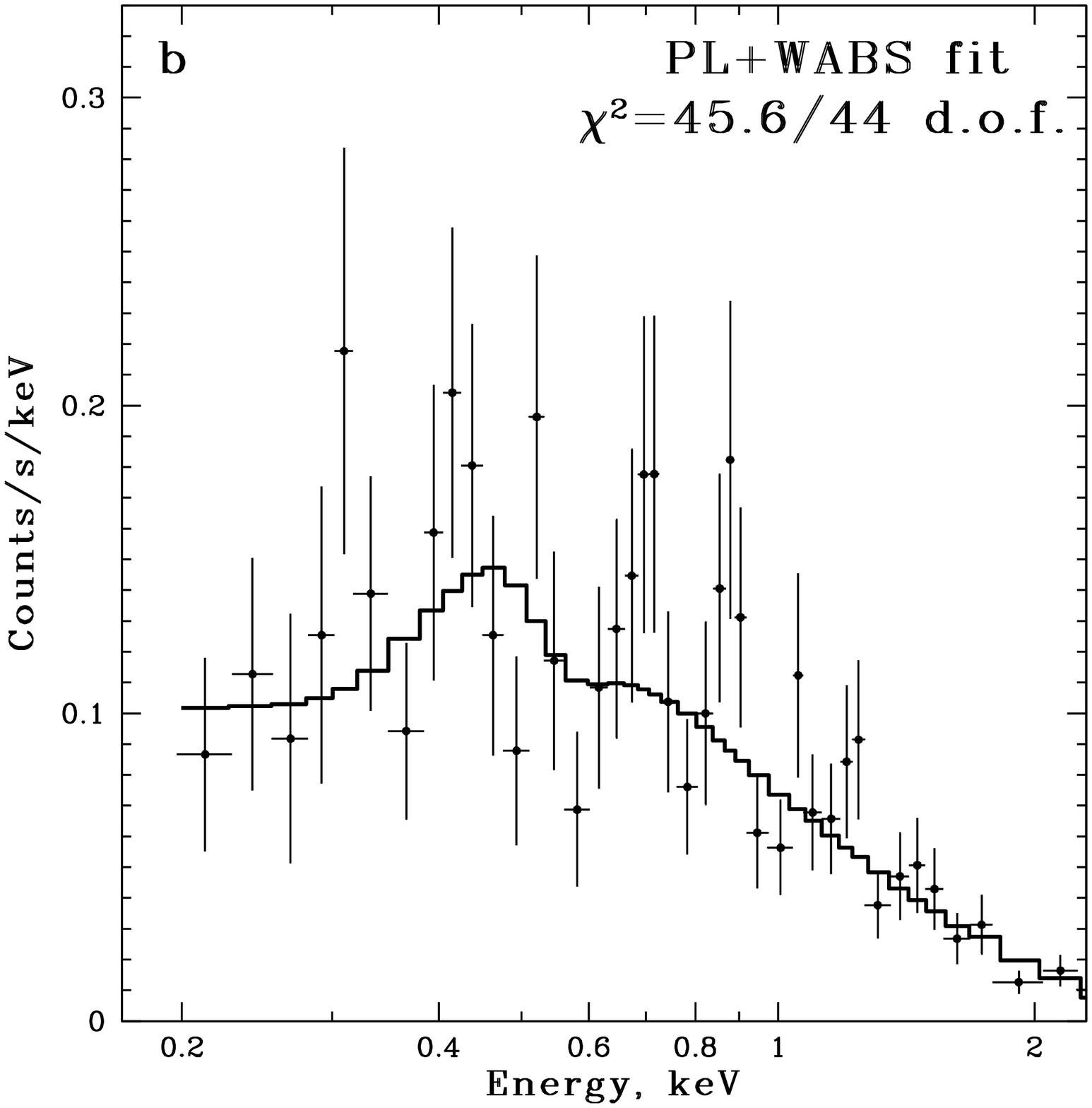}
\end{minipage}
\begin{minipage}{5.7cm}
\epsfxsize=5.9cm
\epsffile{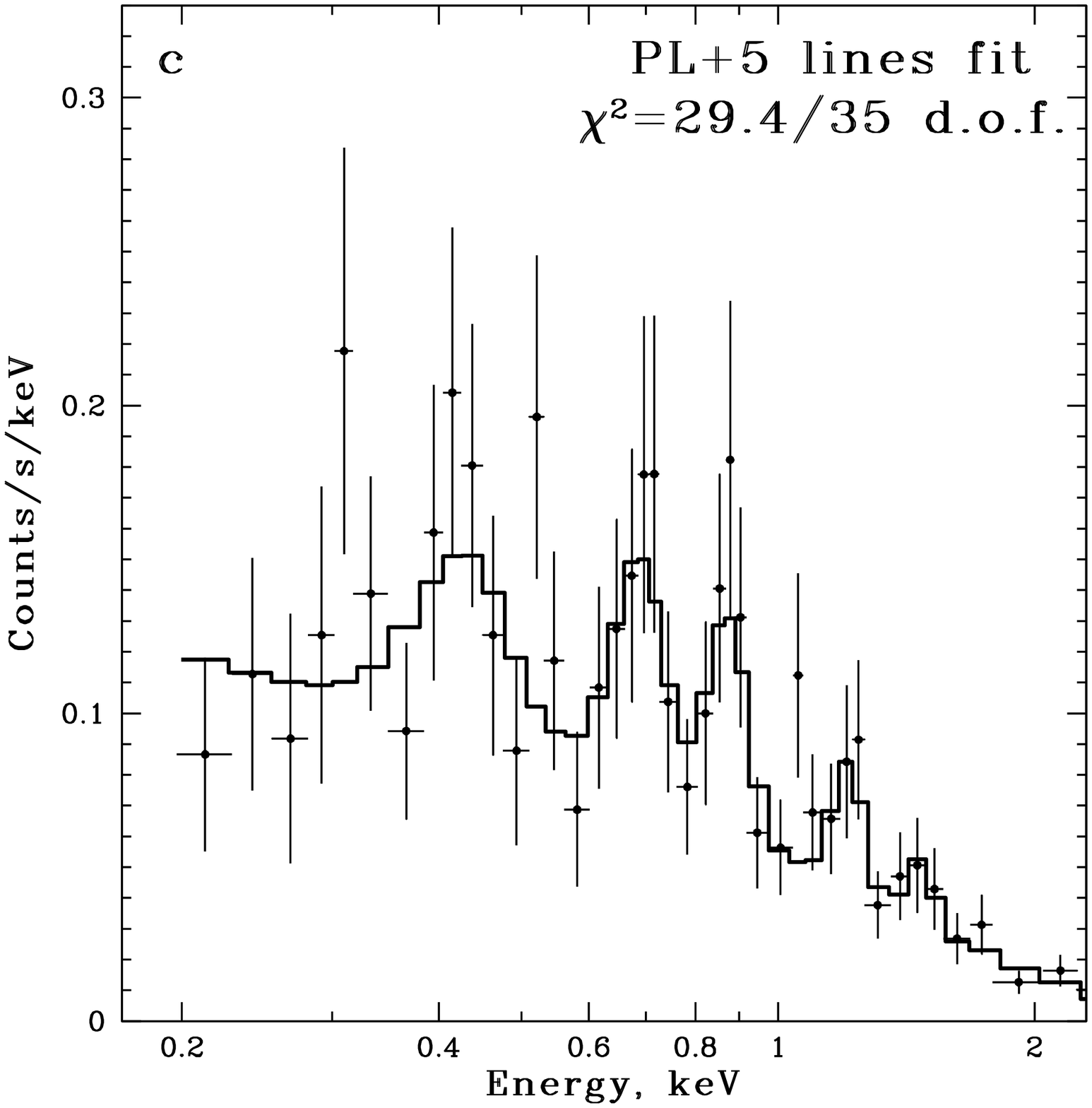}
\end{minipage}
}
\hbox{
\begin{minipage}{5.7cm}
\epsfxsize=5.9cm
\epsffile{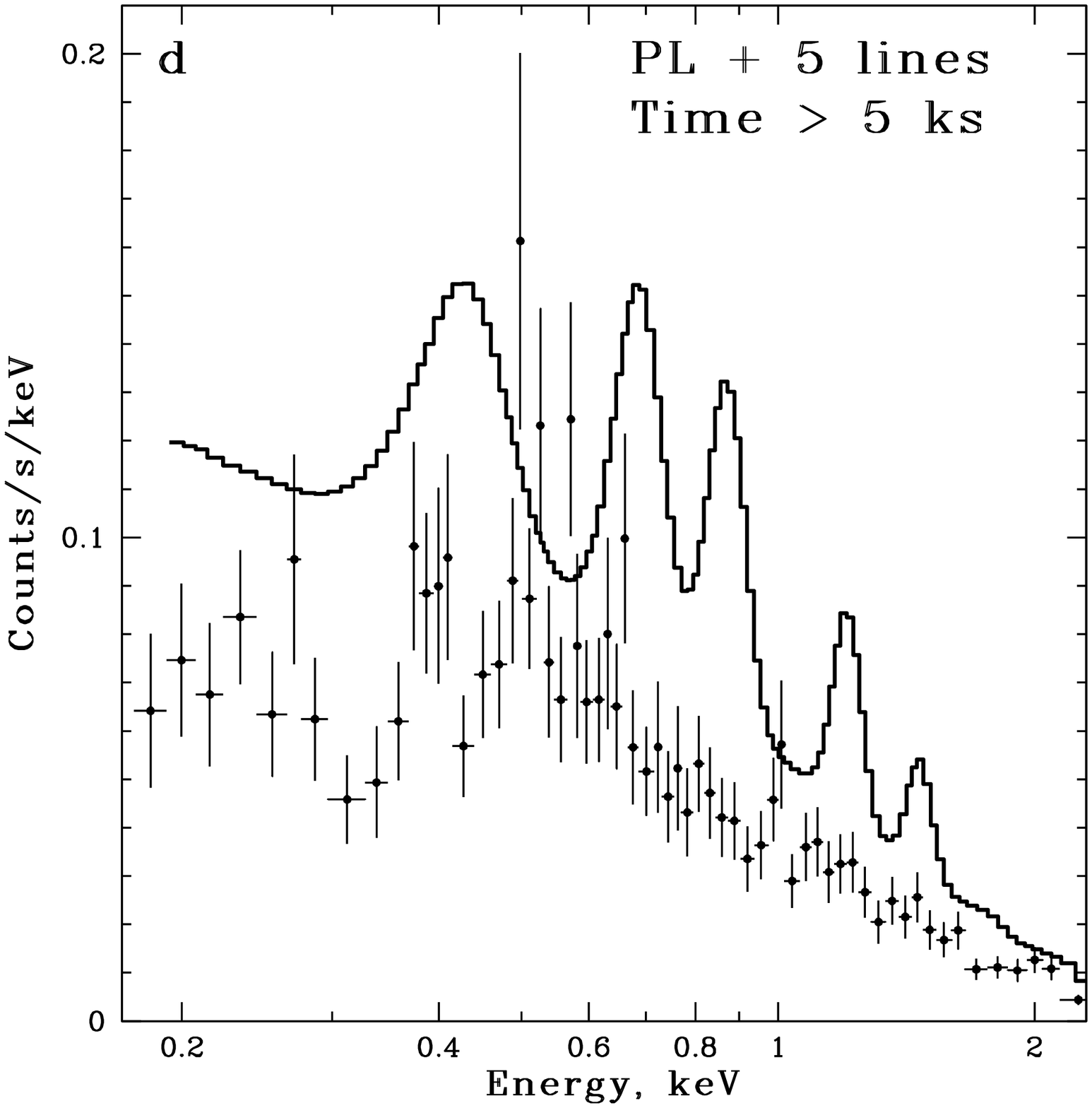}
\end{minipage}
\begin{minipage}{5.7cm}
\epsfxsize=5.9cm
\epsffile{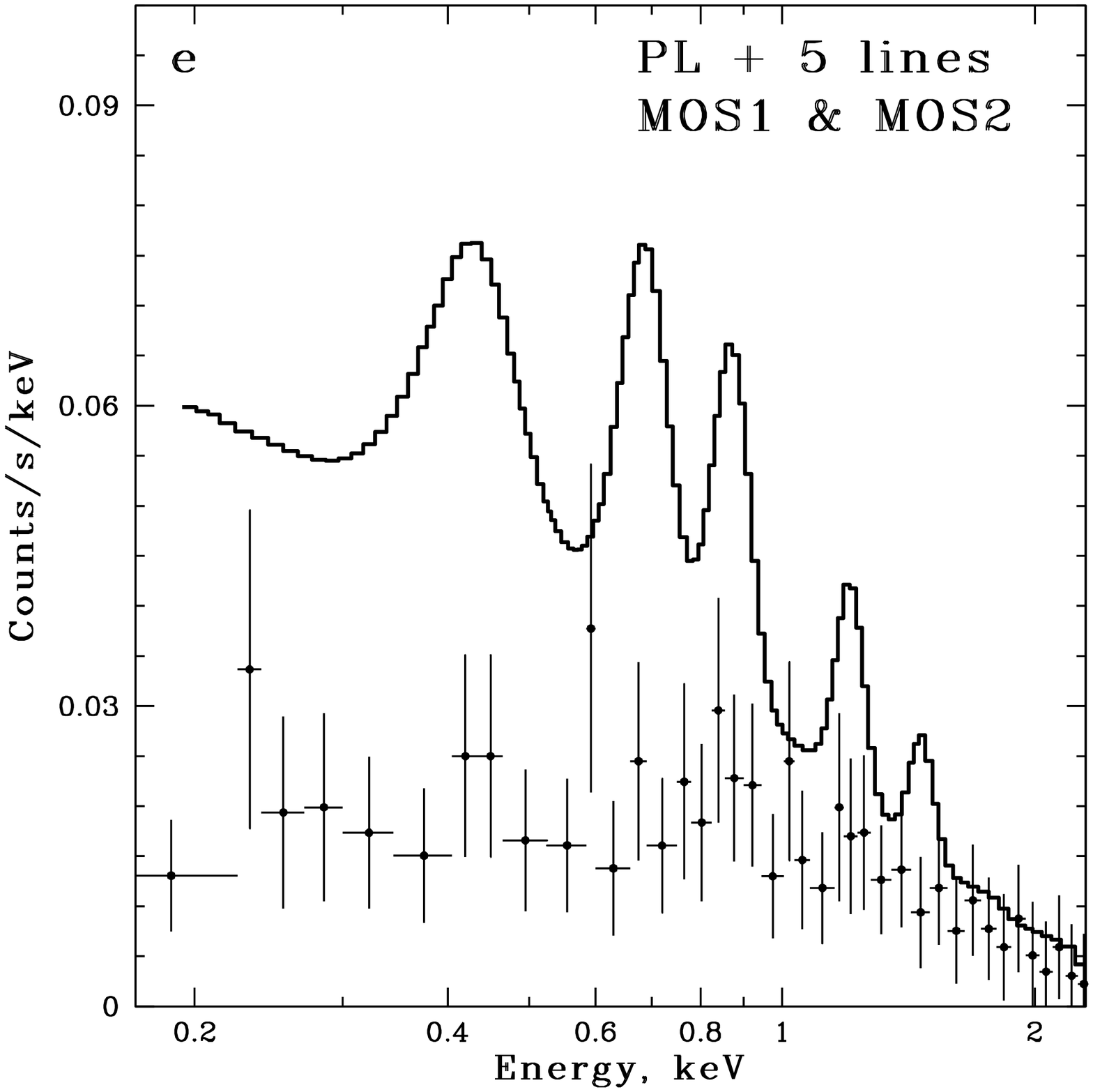}
\end{minipage}
\begin{minipage}{5.7cm}
\epsfxsize=5.9cm
\epsffile{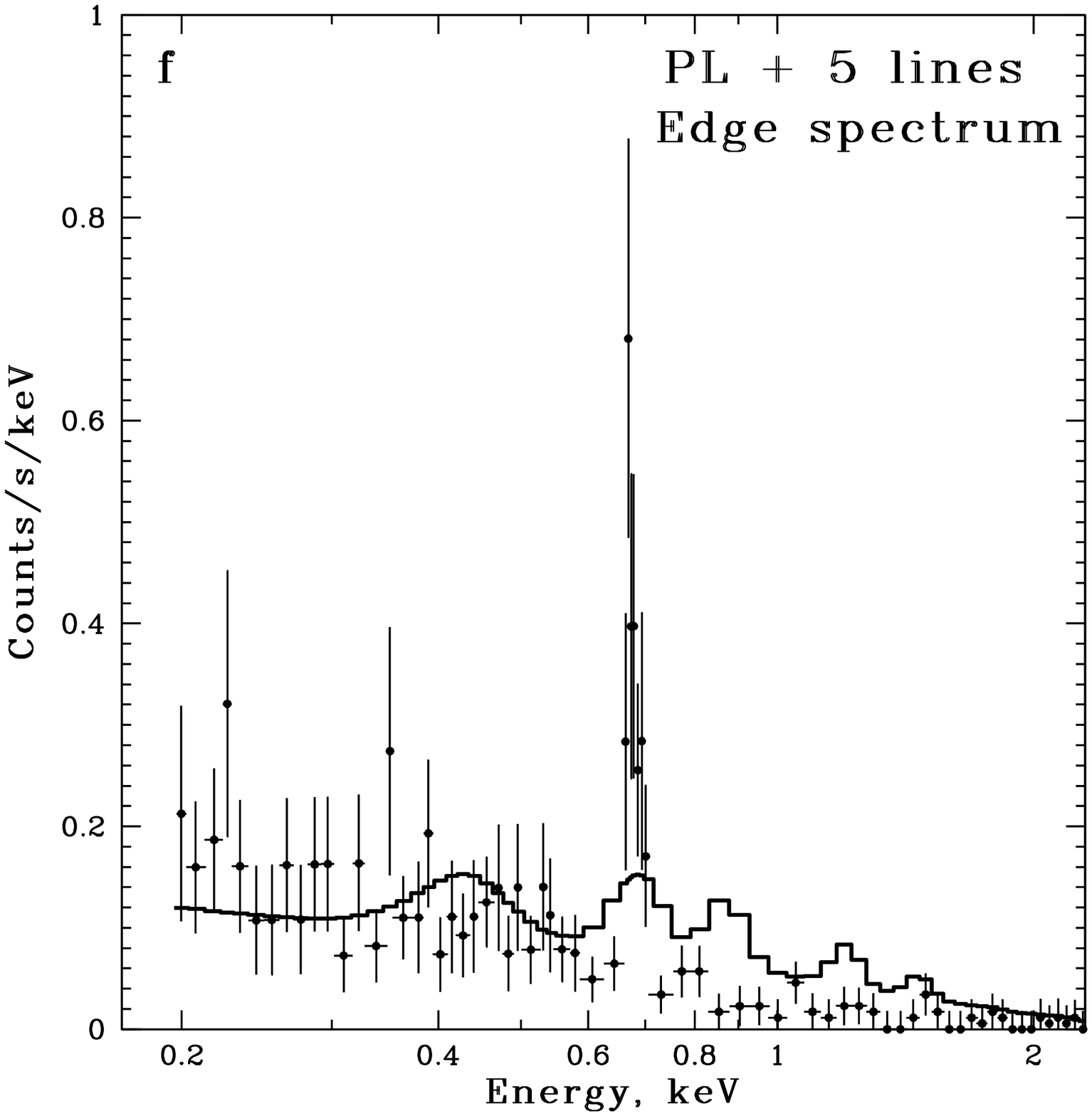}
\end{minipage}
}
}
\caption{\small
Fine structure of the GRB~011211 afterglow spectrum collected by the PN 
detector differs dramatically before and after the {\it XMM-Newton}
satellite reorientation.  PN detector image ({\it Panel a}) shows
two positions of the source on the detector, before (Position 1) and
after (Position 2) the reorientation. The spectrum collected in the Position
1 ({\it Panels b, c}) is more noisy and can be fit either to
simple power-law + absorption model ({\it Panel b}) or to the same model 
with addition of 5 extra emission lines ({\it Panel c}, Reeves et al. 2002a).
The spectrum obtained in the Position 2 do not show any evidence for
these ``lines'' ({\it Panel d}: {\it solid histogram} is the same model 
as on {\it Panel c}, {\it crosses} represent the PN spectrum after the 
reorientation). Data of MOS1 and MOS2 detectors, where the source was close 
to the CCD center (see Fig.1a) also do not show any lines
({\it Panel e}: {\it solid histogram} is the model from 
{\it Panel c}, {\it crosses} represent the sum of MOS1 and MOS2 counts
collected during the first 5 ks of the observation, simultaneously with
the PN counts shown in {\it Panels b, c}). {\it Panel f} shows 
strong dominating ``feature'' in the background spectrum collected over
the PN CCD boundary ({\it crosses}), which coincides with the 0.7~keV 
``line'' seen in the spectrum of GRB~011211 in Position 1 
(this spectrum is represented by the {\it solid histogram}, 
as in previous panels).
}\label{images}
\end{figure}


\begin{references}{}
\reference{Altieri} Altieri, B., Schartel, N., Santos, M., Tomas, L., 
Guainazzi, M., Piro, L., Parmar, A. 2000, GCN Circ. 869
\reference{Antonelli} Antonelli, A., Piro, L., Vietry, M., et al. 2000, ApJ, 
545, L39
\reference{Burud} Burud, I., Rhoads, J., Fruchter A., \& Hjorth J., on behalf 
of GRACE (Gamma-Ray Afterglow Collaboration at Eso), 2001, GCN Circ. 1213
\reference{Costa} Costa, E., Frontera, F., Heise, J., et al. 1997,
{\it Nature}, 387, 783
\reference{Covino} Covino, S., Ghisellini, G., Saracco, P., et al., 2002,
GCN Circ. 1214
\reference{Dickey \& Lockman} Dickey, J.\ M., \& Lockman, F.\ J. 1990, 
ARAA, 28, 215 
\reference{Frontera} Frontera, F., Amati, L., Guidorzi, C., Montanari, E.,
Costa, E., Feroci, M., Piro, L., Heise, J., in't Zand, J.J.M., 2002,
GCN Circ. 1215
\reference{Fruchter} Fruchter, A., Vreeswijk, P., Rhoads, J., Burud, I., 
2001, GCN Circ. 1200
\reference{Fynbo} Fynbo, J.\ P.\ U., Moller, P., Milvang-Jensen, B., 
Burud, I., Andersen, M.\ I., Pedersen, H., Jensen, B.\ L., Hjorth, J., 
Gorosabel, J. 2000, GCN Circ. 867
\reference{Gandolfi} Gandolfi, G., on behalf of BeppoSAX Mission 
Scientist, 2001, GCN Circ. 1189
\reference{Gladders} Gladders, M., Holland, S., Garnavich, P. M., Jha, S., 
Stanek, K. Z., Bersier, D., Barrientos, L.F., 2001, GCN Circ. 1209
\reference{GranotSari} Granot, J., Sari, R. 2002, ApJ, 568, 820
\reference{Grav} Grav, T., Hansen, M.W., Pedersen, H., Hjorth, J.,
Michelsen, R., Jensen, B.L., Andersen, M.I., Gorosabel, J., Fynbo, J.U., 
2001, GCN Circ. 1191
\reference{Harrison} Harrison, F.A., Yost, S.A., Sari, R., 
Berger, E., Galama, T., et al. 2001, ApJ, 559, 123
\reference{Zand} in't Zand, J.J.M., Kuiper, L.M., Amati, L., Antonelli, L.A.,
Butler, R.C., et al. 2001, ApJ, 559, 710
\reference{Odell} Odell, S.\ L., Puschell, J.\ J., Stein, W.\ A., Owen, F., 
Porcas, R.\ W., Mufson, S., Moffett, T.\ J., Ulrich, M.-H. 1978, ApJ, 224, 22
\reference{Piran} Piran, T. 1999, Phys.Rep., 314, 575
\reference{Enrico} Ramirez-Ruiz, E., Trentham, N., \& Blain, A.~W. 2002,
M.N.R.A.S., 329, 465
\reference{Reeves2000} Reeves, J.\ N., \& Turner, M.\ J.\ L. 2000, MNRAS, 
316, 234
\reference{Reeves2002} Reeves, J.\ N., Watson, D., Osborne, J.\ P., 
Pounds, K.\ A., O'Brien, P.\ T. et al., 2002a, {\it Nature}, 416, L512 (RWO)
\reference{Reeves2002} Reeves, J.\ N., Watson, D., Osborne, J.\ P., 
Pounds, K.\ A., O'Brien, P.\ T. 2002b, astro-ph/0206480
\reference{Santos} Santos-Lleo, M., Loiseau, N., Rodriguez, P., Altieri, B.,
and Schartel, N., 2001, GCN Circ. 1192
\reference{SariEsin} Sari, R., Esin, A.A. 2001, ApJ, 548, 787
\reference{Simon} Simon, V., Hudec, R., Pizzichini, G., Masetti, N.,
2001, GCN Circ. 1211
\reference{Smith} Smith, D.\ A., Levine, A.\ M., Remillard, R., Hurley, K., 
Cline, T. 2000, GCN Circ. 861
\reference{Strueder01}Strueder, L. et al., 2001, A\&A, L18
\reference{Turner01}Turner, M. et al., 2001, A\&A, 365, L27
\end{references}
\end{document}